\documentclass[runningheads]{llncs}
\usepackage{graphicx}
\pdfoutput=1
\usepackage{amsmath}
\usepackage{algorithmic}
\usepackage{array}

\usepackage{subfigure}

\usepackage[nocompress]{cite}

\begin{document}
\title{Differentially Private Streaming Data Release under Temporal Correlations via Post-processing}
\titlerunning{DP Streaming Data Release under  Correlations via Post-processing}
%
\author{Xuyang Cao\inst{1} \and
Yang Cao\inst{1} \and
Primal Pappachan\inst{2} \and
Atsuyoshi Nakamura\inst{1} \and
Masatoshi Yoshikawa\inst{3} }
\authorrunning{Xuyang C. et al.}
%
\institute{Graduate School of IST, Hokkaido University, Japan \and
Portland State University, USA \and
Faculty of Data Science, Osaka Seikei University, Japan}
\maketitle              
\begin{abstract}
The release of differentially private streaming data has been extensively studied, yet striking a good balance between privacy and utility on temporally correlated data in the stream remains an open problem. 
Existing works focus on enhancing privacy when applying differential privacy to correlated data, highlighting that differential privacy may suffer from additional privacy leakage under correlations; consequently, a small privacy budget has to be used which worsens the utility. 
In this work, we propose a post-processing framework to improve the utility of differential privacy data release under temporal correlations. We model the problem as a maximum posterior estimation given the released differentially private data and correlation model and transform it into nonlinear constrained programming. Our experiments on synthetic datasets show that the proposed approach significantly improves the utility and accuracy of differentially private data by nearly a hundred times in terms of mean square error when a strict privacy budget is given. 

\keywords{Differential Privacy  \and Data Correlations \and Time-series Stream \and Continual Data Release \and Post-processing.}
\end{abstract}
\section{Introduction}

\begin{figure}[t]
    \centering
    \includegraphics[width= 0.9\linewidth]{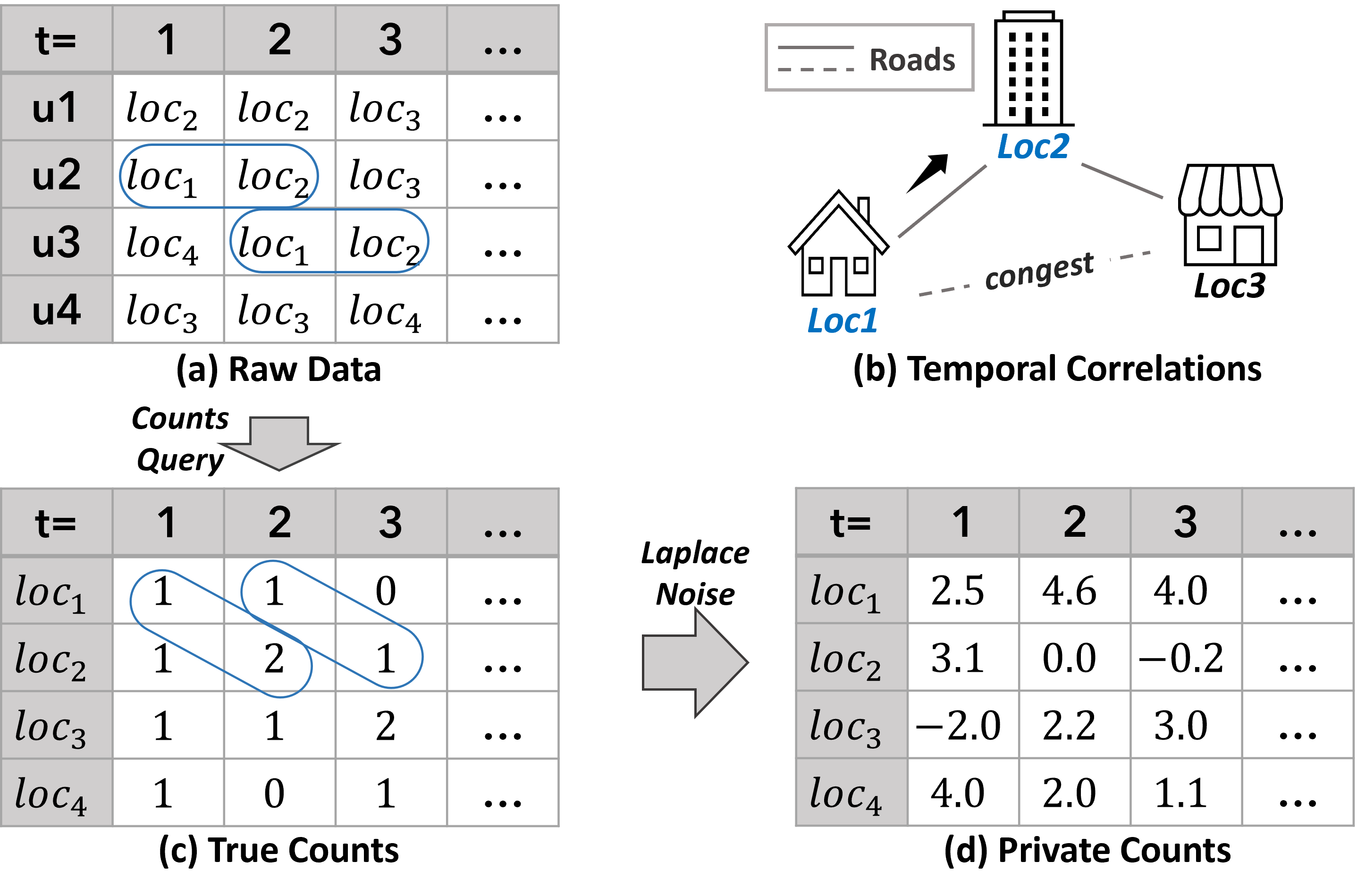}
    \caption{Scenario: Differentially Private Streaming Data Release.}
    \label{fig:problem}
\end{figure}

Data collection and analysis in many real-world scenarios are performed in a streaming fashion, such as location traces \cite{GoogleMaps19}, web page click data\cite{erlingssonRAPPORRandomizedAggregatable2014}, and real-time stock trades. However, releasing data continuously may result in privacy risks.
To this end, \textit{differentially private streaming data release} have been thoroughly studied  \cite{cao_differentially_2016,cao_differentially_2015, chenPeGaSusDataAdaptiveDifferentially2017, dworkDifferentialPrivacyContinual2010, erlingssonRAPPORRandomizedAggregatable2014, fanFASTDifferentiallyPrivate2013, friedmanPrivacyPreservingDistributedStream2014, georgioskellarisDifferentiallyPrivateEvent2014, mirPanprivateAlgorithmsStatistics2011,Cunningham_2021}.
The curator of the database can use a differentially private mechanism, such as Laplace Mechanism (LM), that adds noises to the query results at each time point for satisfying a formal privacy guarantee called $\epsilon$-Differential Privacy ($\epsilon$-DP) \cite{dwork2008differential}, where $\epsilon$ is the parameter (i.e., privacy budget) controlling trade-off between privacy protection and utility of data release.
A small $\epsilon$ indicates a high level of privacy and thus requires adding a larger amount of noise. Taking location traces as an example to elaborate, Fig.~\ref{fig:problem} (a) (c) (d) illustrate how differentially private location statistics are released using LM at each time point where (a) represent real-time location raw data sets (i.e., values of longitude and latitude of residence, company, shopping mall respectively) of users in a database $D$ collected by devices with GPS sensors (i.e., GPS, GNSS) \cite{GoogleMaps19}, (c) are the true counts of each location computed by a count query function $f(D)$ and (d) what will be released and sent to the public are streaming private counts through a differentially private mechanism such as Laplace Mechanism (LM) \cite{Dwork06}.

However, recent studies \cite{Cao18, caoQuantifyingDifferentialPrivacy2017, Zhu15, yang2015bayesian, Song17, kiferPufferfishFrameworkMathematical2014} reveal that, when the data are correlated, more noises have to be added to prevent leakages which deteriorates the utility. They point out that differential privacy algorithms suffer extra privacy leakage on correlated data and develop techniques to enhance differential privacy with a smaller $\epsilon$.
In the context of streaming data release, a Markov chain could be used to model the temporal correlations. 
For example, as shown in Fig. ~\ref{fig:problem} (b), temporal correlation is manifested as the transition probabilities between different locations, which can be obtained through public information such as road networks or traffic data. Based on the temporal correlation presented in Figure 1(b), we have the probability of users proceeding from location $loc1$ to $loc2$ will be $Pr(l^{t+1}=loc2|l^{t}=loc1)=1$ if we have the knowledge that another road is congested. Cao et al. \cite{Cao18, caoQuantifyingDifferentialPrivacy2017} quantified such a private leakage and proposed a special privacy definition on temporal correlated data named $\alpha\text{-}DP_\mathcal{T}$, to calibrate a smaller privacy budget in order to cover the extra privacy leakage caused by temporal correlations. 
Song et al. \cite{Song17} proposed Wasserstein Mechanism for Pufferfish privacy (i.e., a privacy notion that generalizes differential privacy) and Markov Quilt Mechanism specifically when correlation between data is described by a Bayesian Network or a Markov chain.
Similar to \cite{Cao18,caoQuantifyingDifferentialPrivacy2017}, they calculate an enlarged $\epsilon$ to enhance the privacy but sacrifice more utility. 
Hence, the challenge is how to boost the utility of differentially private streaming data release on temporally correlated data.

Our approach to addressing the aforementioned issue involves capitalizing on the existing temporal correlations as prior knowledge about the original data through post-processing. Although \textit{post-processing} \cite{hay2010, Lee15, mckenna19, wang21} has been extensively researched as a means to enhance the utility of differential privacy, current methods are ill-equipped to deal with temporal correlations. Post-processing primarily aims to refine differentially private (noisy) results by enforcing them to comply with certain ground-truth constraints or prior knowledge about the data.
For instance, \textit{deterministic} consistency constraints between data points are frequently employed in previous studies to represent inherent properties of the data (e.g., released counts in histograms should be integers). In this study, we apply the post-processing technique to improve the utility of differentially private streaming data release in the presence of temporal correlations. 
By accounting for temporal correlations along with other consistency constraints, we strive to obtain the most accurate current counts which could be estimated from previous private counts while approximating the true current counts.

\begin{figure}[t]
    \centering
    \includegraphics[width= \linewidth]{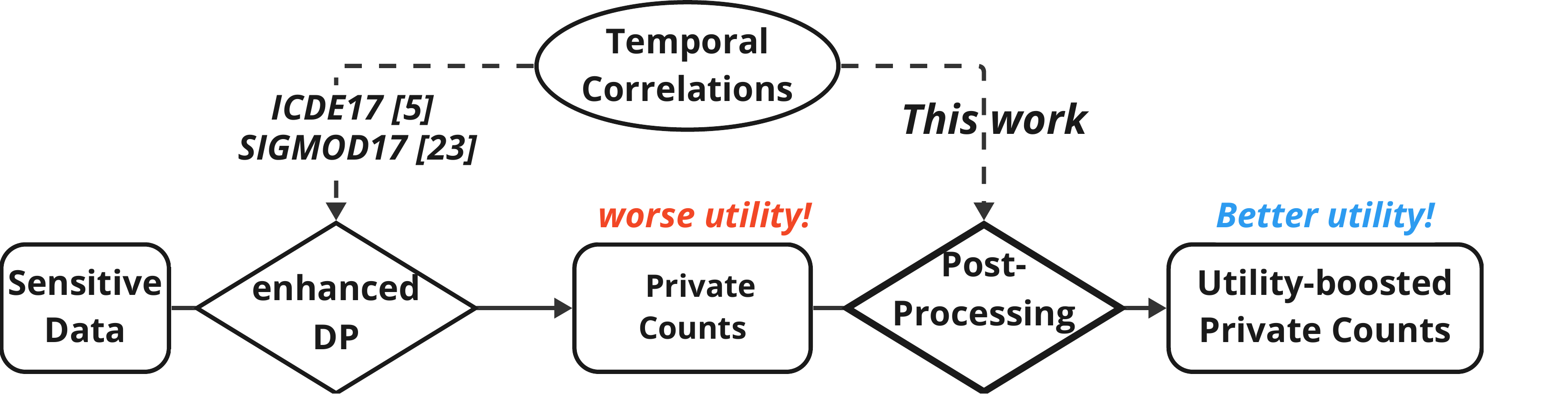}
    \caption{Existing studies \cite{caoQuantifyingDifferentialPrivacy2017, Song17} propose approaches for enhancing DP on temporally correlated data; however, these methods sacrifice utility. This work tackles this problem by utilizing temporal correlations as prior knowledge about the data for post-processing purposes.}
    \label{fig:MAP}
\end{figure}

In this study, we formulate post-processing as a nonlinear optimization problem within the Maximum A Posteriori (MAP) framework, accounting for both probabilistic constraints of temporal correlations and deterministic consistency constraints. 
Similar to \cite{Cao18, caoQuantifyingDifferentialPrivacy2017, Song17}, we assume that temporal correlations are public knowledge and are expressed by a transition matrix. 
As illustrated in Fig. \ref{fig:MAP}, our approach leverages the transition matrix to enhance the utility of differentially private counts.
Thus, we pose the problem of determining the most plausible counts that satisfy the constraints (both probabilistic and deterministic) and exhibit the least distance from the released private data. 
To model this probabilistic distribution, we employ the knowledge of Laplace noise distribution and introduce a Markov chain model to calculate the distribution of true counts. Finally, extensive experiments demonstrate and validate the effectiveness of our methods.

To summarize, our contributions are as follows:
\begin{itemize}
\item To the best of our knowledge, this paper presents the first attempt to enhance the utility of differentially private data release under temporal correlations. We propose a post-processing framework using maximum a posteriori estimation, which incorporates both probabilistic correlations and deterministic constraints.
\item We implement the post-processing framework for temporal correlations in the differentially private continual data release. Specifically, we formulate this problem as constrained nonlinear programming, which can be solved using off-the-shelf optimization software.
\item Our experiments on synthetic data demonstrate the effectiveness of the proposed approach. We show that the utility of differentially private data is significantly improved, with nearly a $100$-fold reduction in mean square error under a strict privacy budget, while preserving temporal correlations between data.
\end{itemize}

We would like to note that this work is an extension of our previous poster paper \cite{cao22poster}, in which we briefly presented the idea without delving into technical details. This paper provides a clearer and more in-depth exploration of our previous work, offering a comprehensive understanding of the proposed MAP-based post-processing framework.

\section{Related Work}
 
Several well-studied methods exist to enhance the utility of differentially private data, as post-processing is an effective tool. In this section, we review related works on improving the utility of private data through post-processing.

One of the most widely studied approaches for utility enhancement is the utilization of \emph{consistency constraints} \cite{hay2010, Lee15} in data (e.g., the \textit{sum} of the released data should be a fixed number, or the released values should be \textit{integer} in the case of counting queries). In our location traces scenario, these consistency constraints can be expressed by the sum of location records or the total number of users as a fixed value (e.g., $n$) for each time point, with counts always being integers.
Previous works formulate the problem as a least squares estimation (LSE) problem \cite{hay2010} or a maximum likelihood estimation (MLE) problem \cite{Lee15}, demonstrating the effectiveness of such post-processing approaches.

Hay et al. \cite{hay2010} focused on improving the accuracy of private histograms through post-processing, solving an LSE problem given consistency constraints such as $sum$, $sorted$, and $positive$ to find the 'closest' private histograms that also satisfy these constraints.
Furthermore, Lee et al. \cite{Lee15} considered noise distribution (Laplace distribution in their scenario) to boost the utility of private query results. They formulated their post-processing problem as an MSE problem and employed the ADMM algorithm to solve the programming problem.

However, when publishing statistics continually, the data points are often temporally correlated. The post-processing methods mentioned above only focus on single-time data release and cannot efficiently capture probabilistic temporal correlations.
Moreover, it remains unclear how to formulate \textit{probabilistic} correlations as constraints, as existing works assume \textit{deterministic} constraints as prior knowledge about the data.
We also observe that many existing works on differentially private streaming data release neither provide a formal privacy guarantee under temporal correlations \cite{cao_differentially_2016,cao_differentially_2015, chenPeGaSusDataAdaptiveDifferentially2017, dworkDifferentialPrivacyContinual2010, erlingssonRAPPORRandomizedAggregatable2014, fanFASTDifferentiallyPrivate2013, friedmanPrivacyPreservingDistributedStream2014, georgioskellarisDifferentiallyPrivateEvent2014, mirPanprivateAlgorithmsStatistics2011} nor offer reasonable utility for private outputs.
Therefore, our study represents the first attempt to enhance the utility of DP with formal privacy under temporal correlations.

\section{Preliminaries}

\subsection{Differential Privacy}
Informally, the DP notion requires any single element in a dataset to have only a limited impact on the output. Namely, if $D$ and $D^{\prime}$ are two \emph{neighboring} databases, the difference in outputs of executing a randomized algorithm on these databases should be minimal\cite{DPBook16}.
\begin{definition}
    ($\epsilon$-DP) A randomized mechanism $\mathcal{M}$ is said to satisfy $\epsilon$-DP, where $\epsilon \ge 0$, if and only if for any neighboring datasets $D$ and $D^{\prime}$ that differ on one element, we have
    \begin{equation*}
    \forall T \subseteq Range(\mathcal{M}): Pr(\mathcal{M}(D)\in T) \leq e^{\epsilon}Pr(\mathcal{M}(D^{\prime})\in T)
    \end{equation*}
    where $Range(\mathcal{M})$ represents the set of all possible outputs of the algorithm of mechanism $\mathcal{M}$, the parameter $\epsilon$ represents the privacy budget.
\end{definition}

\subsection{The Laplace Mechanism}
The Laplace Mechanism\cite{Dwork06} is the first and probably most widely used mechanism for DP. It satisfies $\epsilon$-DP by adding noise to the output of a numerical function\cite{DPBook16}.
\begin{definition}
    (Global sensitivity) Let $D\approx D^{\prime}$ denote that $D$ and $D^{\prime}$ are neighboring. The global sensitivity of a query function $f$, denoted by $\Delta$, is given below
    \begin{equation*}
        \Delta = \max_{D\approx D^{\prime}}|f(D)-f(D^{\prime})|
    \end{equation*}
\end{definition}
According to the definition of DP, the probability density function of the noise should have the property that if one moves no more than $\Delta$ units, the probability should increase or decrease no more than $e^{\epsilon}$. The distribution of noise that naturally satisfies this requirement is $Lap(\frac{\Delta}{\epsilon})$\cite{DPBook16}, which denotes a Laplace distribution with location 0 and scale $\frac{\Delta}{\epsilon}$.
\begin{theorem}
    (Laplace Mechanism, LM) For any function $f$, the Laplace mechanism $A_{f}$ that adds i.i.d noise to each function output $f$ satisfies $\epsilon$-DP.
    \begin{equation*}
        A_{f}(D) = f(D) + Lap\left(\frac{\Delta}{\epsilon}\right).
    \end{equation*}
\end{theorem}

Commonly, we denote the scale parameter using $\lambda = \frac{\Delta}{\epsilon}$.

\section{Problem Statement}


This section will introduce and formulate the primary issue we aim to address. 
First, below we present the notations used throughout this paper. 
We use ${D}$ to represent a \emph{bounded database} consisting of $n$ users. 
We prefer to use bold letters to indicate vectors. 
We use $r^{t}_{l}$ to denote a specific query output at a given time point $t$ and location $l$. 
$\mathcal{T}$ represents a transition matrix modeling temporal correlations between data. 
More detailed notations are in Table ~\ref{tab:notations}.

\begin{table}[t]
    \centering
    \renewcommand{\arraystretch}{1.5}
    \caption{Notations}
    \label{tab:notations}
    \begin{tabular}{|c|p{7cm}|}
        \hline
        $D$ & A bounded database \\ \hline
        $\mathbf{Loc}$ & Value domain of locations of all users \\ \hline
        $l^{t}_{i}$ & The location information of $user_i$ at time t, $user_i \in U$, $l^{t}_{i} \in \mathbf{Loc}$ \\ \hline
        $\mathcal{M}$ & A differential privacy mechanism over D \\  \hline
        $\mathbf{R}$ & The set of real continual time-series query outputs \\ \hline
        $\tilde{\mathbf{R}}$ & The set of added-noise continual time-series query outputs \\ \hline
        $R^{t}$ & The set of query outputs at time $t$, $R^{t} \subseteq \mathbf{R}$ \\ \hline
        $r^{t}_{l}$ & A specific query output at time $t$ and location $l$, $l \in \mathbf{Loc}$, $r^{t}_{l} \in R^{t}$ \\ \hline
        $\mathcal{T}$ & The transition matrix of locations \\ \hline
        $\mathbf{P^{t}}$ & The possibility of locations for a single user at time $t$ \\ \hline
        $Pr(\mathbf{\hat{R}})$ & The joint distribution of possible private counts \\ \hline
    \end{tabular}
\end{table}

\textit{\textbf{Temporally Correlated Stream Data.}}
In our scenario of location traces mentioned above, we assume that $n$ people (labeled from $1$ to $n$) staying at $m$ locations (labeled from $1$ to $m$) respectively at single time point $t$ (shown in Fig.~\ref{fig:problem} (a)). Let $\mathbf{Loc}$ denote the sets of locations. 
Naturally, the data at each time point are temporally correlated: for each user, her current location depends on the previous location in the form a transition matrix $\mathcal{T}$.
Without loss of generality, we assume the transition matrix is the same for all users and is given in advance since it can be learned from public information such as road networks.
This assumption follows existing works \cite{caoQuantifyingDifferentialPrivacy2017, Song17}.
\begin{table}[ht]
    \centering
    \caption{Transition Matrix}
    \label{TM}
    \begin{tabular}{|c|c|c|c|}
    \hline
         $Pr(l^{t+1}|l^{t})$ & {\itshape Loc1} & {\itshape Loc2} & {\itshape Loc3}\\ \hline
         {\itshape Loc1} & 0.33 & 0.33 & 0.34\\ \hline
         {\itshape Loc2} & 0.80 & 0.10 & 0.10\\ \hline
         {\itshape Loc3} & 0.05 & 0.90 & 0.05\\ \hline
    \end{tabular}
\end{table}



\textit{\textbf{Differentially Private Stream Data Release.}} A server collects users' real-time locations $l^{t}$ at time $t$ in a database $D$, and aims to release differentially private query results over $D$.
In particular, we consider a query function $f: D\rightarrow \mathbf{N}^{m}$ that counts the total number of people at each location over the entire publishing time $T$, denoted as $f(D)$. The query outputs are represented by $\mathbf{R} = (R^{1},\dots,R^{t} ,\dots,R^{T})$ and $R^{t} = (r^{t}_{1},\dots ,r^{t}_{m})$.
Many existing works, such as \cite{cao_differentially_2016,cao_differentially_2015, chenPeGaSusDataAdaptiveDifferentially2017, dworkDifferentialPrivacyContinual2010, erlingssonRAPPORRandomizedAggregatable2014, fanFASTDifferentiallyPrivate2013, friedmanPrivacyPreservingDistributedStream2014, georgioskellarisDifferentiallyPrivateEvent2014, mirPanprivateAlgorithmsStatistics2011}, have considered a similar problem setting as ours.
However, due to temporal correlations, increased noise is added to the true answers to preserve strict privacy \cite{caoQuantifyingDifferentialPrivacy2017, Song17}, which reduces the utility of the released private counts. 
Our question is: {\textit{can we leverage the temporal correlations to improve the utility of differentially private data via post-processing}} (while preserving the enhanced privacy as \cite{caoQuantifyingDifferentialPrivacy2017, Song17})?


\section{Methodology}
In this section, we will explain how to formulate the post-processing problem for streaming data release under temporal correlations.
To address the above-mentioned challenge, we use post-processing, allowing us to refine the private counts using publicly known prior knowledge. 

\textbf{Intuition}. Our core idea is that the temporal correlations can be seen as probabilistic constraints on the data.
We can formulate the problem as determining the most probable query outputs $\mathbf{\hat{R}}$ that satisfy such constraints when given $\mathbf{\tilde{R}}$, leveraging the knowledge of $\mathcal{T}$ as shown in Fig. \ref{fig:problem} (b). Specifically, we aim to solve the programming problem of maximizing $Pr(\mathbf{\hat{R}}|\mathbf{\tilde{R}})$, subject to the \emph{transition matrix} and other \emph{consistency constraints}.
Our method will demonstrate that the estimation depends on the noise distribution and the joint distribution of true counts, which are determined by the mechanism used and the inherent correlations within the raw data.

\subsection{Maximum A Posterior Estimation Framework for Correlated Data}
Firstly, we propose a Maximum A Posterior (MAP) Estimation framework to assist formulating probabilistic post-processing problem.

\begin{definition}
    (MAP Framework) Let $D$ be a bounded database with $n$ records. A post-processing approach is feasible under a framework $\mathcal{F}(\mathcal{M,C})$ if for all noisy query results $\tilde{Q} \in \mathcal{O}$ through a given privacy mechanism $\mathcal{M}$, we have
    \begin{align}
        P(\hat{Q}|\tilde{Q}) &= \frac{P(\tilde{Q}|\hat{Q})P(\hat{Q})}{P(\tilde{Q})}, \\
    \hat{Q}^{*} &= \arg \max_{\hat{Q}}P(\tilde{Q}|\hat{Q})P(\hat{Q})
    \end{align}
     where $\mathcal{C}$ represents correlations between data for all true query results $Q$, $\mathcal{O}$ is denoted as all possible output set of $\mathcal{M}(D)$, $\hat{Q}$ and $\hat{Q}^{*}$ are variable and our desired `closest' query result which also meets correlation $\mathcal{C}$ respectively.
\end{definition}

We apply MAP Framework to solve post-processing problem of streaming data release under temporal correlations. 
Given a mechanism $\mathcal{M}$(i.e., Laplace Mechanism here) and temporal correlations $\mathcal{C}$ between data, the `closest' private counts $\mathbf{\hat{R}}$ is tended to be obtained by calculating the maximum of the posterior possibility under MAP framework $\mathcal{F}(\mathcal{M,C})$
\begin{equation}
    \label{original equation}
    Pr(\mathbf{\hat{R}}|\mathbf{\tilde{R}}) = \frac{Pr(\mathbf{\tilde{R}}|\mathbf{\hat{R}})Pr(\mathbf{\hat{R}})}{Pr(\mathbf{\tilde{R}})}
\end{equation}
subjecting to the correlations $\mathcal{C}$ and other constraints if exist. 
For convenience, the logarithm form of the above formula is applied
\begin{equation}
    \ln Pr(\mathbf{\hat{R}}|\mathbf{\tilde{R}}) = \ln Pr(\mathbf{\tilde{R}}|\mathbf{\hat{R}})+ \ln Pr(\mathbf{\hat{R}})- \ln Pr(\mathbf{\tilde{R}})
    \label{log original equation}
\end{equation}

Therefore, the objective `closest' query outputs (achieve the maximum of \eqref{original equation}) after post-processing will be
\begin{equation}
    \label{objective equation}
    \mathbf{\hat{R}}^{*} = \arg \max_{\mathbf{\hat{R}}} 
    \{ \ln Pr(\mathbf{\tilde{R}}|\mathbf{\hat{R}})+ \ln Pr(\mathbf{\hat{R}}) \}
\end{equation}
when the private counts $\mathbf{\tilde{R}}$ is given.

In essence, the first term and the second term of right side of \eqref{objective equation} come from $\mathcal{M}$ and $\mathcal{C}$ respectively. What makes it different from prior works is that we focus on calculating the joint distribution of private counts $Pr(\mathbf{\hat{R}})$ which are simply viewed as a uniform distribution, namely a constant, in most of previous works. We point out that it cannot be omitted when there are correlations between data especially under temporal correlations.

\subsection{Calculation of Terms of Objective Equation}
The next steps are how to calculate the left two terms in the right side of \eqref{objective equation}.

\subsubsection{Calculation of the first term}
For this term, it tells us that noises should be considered while improving accuracy and \cite{Lee15} also points out that we are able to formulate it into a $L_{1}$ function if the noises come from LM\footnote{Please note that our method can be applied to other mechanisms. However, for the duration of this article, we have temporarily chosen to default to the Laplace Mechanism.}.
Thus, we formulate the first term of \eqref{objective equation} in the following
\begin{equation}
    \label{first term}
    \ln Pr(\mathbf{\tilde{R}}|\mathbf{\hat{R}}) = -\frac{1}{\lambda}||\mathbf{\tilde{R}}-\mathbf{\hat{R}}||_{L_{1}} + \textit{Const.}
\end{equation}

\subsubsection{Calculation of the second term}
A \emph{Markov Chain} model is introduced to calculate the possibilities of single user's locations released in continual time-series stream because the possibility of present location only relies on previous one. 
With the transition matrix and a prior distribution of locations of single user at $t=1$, we are able to calculate user's probability distribution of location at any time. 
We introduce two policies to obtain the prior distribution: (a) the first one is to use the normalized frequency of private counts at $t=1$, $\tilde{R}^{1}$ (\emph{frequency p-d}); (b) another is to simply use a uniform distribution (\emph{uniform p-d}) instead.
Consequently, we can derive all the possibilities of moving next locations at each time $t$, $\mathbf{P}^{t}$, expressed as below:
\begin{equation}
    \mathbf{P}^{t} = \mathbf{P}^{t-1}\mathcal{T}
\end{equation}
for each $t \in \{2,\dots,T\} $.
However, a joint distribution of users' locations should be calculated when given a bounded database containing data of $n$ users. 

Note that all of $n$ users are independent here which means their next actions will not be influenced by others. 
With the probability distribution of location of single user at each time, therefore, the joint distribution of all location counts at specific time point can be expressed by a \emph{multinomial distribution}
\begin{equation}
    Pr(R^{t}) = n!\prod_{l}\frac{(\mathbf{P}_{l}^{t})^{r_{l}^{t}}}{r_{l}^{t}!}
\end{equation}
for each $R^{t} \subseteq \mathbf{R}$ where $n$ represents total number of users.

Recall the \emph{Stirling's Approximation}
\begin{equation}
\label{approximation}
    \ln{x!} \approx \frac{\ln 2\pi x}{2}+x\ln\frac{x}{e}
\end{equation}
We apply the approximation \eqref{approximation} to mitigate our calculation
\begin{equation}
    \label{second term}
    \ln Pr(R^{t}) \approx \ln n!+\sum_{l}(r_{l}^{t} \ln\mathbf{P}_{l}^{t}- \frac{\ln2\pi r_{l}^{t}}{2} - r_{l}^{t}\ln \frac{r_{l}^{t}}{e})
\end{equation}

Naturally, our `closest' query answer $\mathbf{\hat{R}}$ also obeys this multinomial distribution the same as true query answer. 

\subsection{Nonlinear Constrained Programming}
We conclude our method of formulating this post-processing problem under temporal correlations into a nonlinear constrained programming problem. 
By calculating the minimum estimation of $-\ln Pr(\mathbf{\hat{R}}|\mathbf{\tilde{R}})$ and combining with \eqref{first term} \eqref{second term}, we finally transform \eqref{objective equation} into a nonlinear constrained programming as below
\begin{align*}
       \text{Minimize } & \frac{1}{\lambda}||\mathbf{\tilde{R}}-\mathbf{\hat{R}}||_{L_{1}}\\
       & - \sum_{t=1}^{T} \sum_{l}(r_{l}^{t}\ln\mathbf{P}_{l}^{t} -  \frac{\ln2\pi r_{l}^{t}}{2} - r_{l}^{t}\ln \frac{r_{l}^{t}}{e}) \\
       \text{Subject to }
        &\sum r_{l}^{t} = n \text{, for each }t \in \{1,2,\dots,T-1,T\}\\
        &r_{l}^{t} \ge 0 \text{, for each }t \in \{1,2,\dots,T-1,T\}
\end{align*}
where $\hat{\mathbf{R}}=\left((r^1_1,\dots,r^1_m),\dots,(r^T_1,\dots,r^T_m)\right)$.

Then, we point out that this nonlinear constrained programming is solvable. By introducing augmented \emph{Lagrangian} to our objective function (O.F.), there are many convergence results proved in the literature (e.g. \emph{ADMM}\cite{MAL-016}) where we could prove the O.F. will finally converge as dual variables converge. Also, variables $r_{l}^{t}$ must satisfy $\sum r_{l}^{t} = n$ and $r_{l}^{t} \ge 0$ simultaneously. Thus, the boundary of $r_{l}^{t}$ is $n \ge r_{l}^{t} \ge 0$.


\textbf{Asymptotical analysis.}
As shown in derived objective function, there are two terms which represent the contribution from Mechanism applied to true counts and Correlations between true counts respectively. As $\epsilon$ approaches zero, the first term, namely $\frac{1}{\lambda}||\mathbf{\tilde{R}}-\mathbf{\hat{R}}||_{L_{1}}$ will also approach to zero because of coefficient $\lambda$. In other words, the second term
$$- \sum_{t=1}^{T} \sum_{l}(r_{l}^{t}\ln\mathbf{P}_{l}^{t} -  \frac{\ln2\pi r_{l}^{t}}{2} - r_{l}^{t}\ln \frac{r_{l}^{t}}{e})$$
will matter the most to objective function when a stricter privacy budget $\epsilon$ is given. Also, we'd like to analyze what the objective function will perform if a `weak' level correlation is given (note that we will provide a mathematical definition of levels of correlations in our Experiments part) such that probabilities of proceed to the next location from previous ones is a fixed value, namely $\mathbf{P}_{l}^{t}$ is a uniform distribution. Then, the second term is able to be `ignored' and the first term
$$\frac{1}{\lambda}||\mathbf{\tilde{R}}-\mathbf{\hat{R}}||_{L_{1}}$$
will thus matter the most to solutions. We should point out that our framework will result in an MLE problem such that post-processing problem mentioned by  \cite{Lee15} if there is no correlations amount original data.

\section{Experiments}
In this section, we present experimental results that demonstrate the effectiveness of our proposed MAP framework for post-processing continuous data release under temporal correlations. To validate our method, we apply it to both synthetic and real-world datasets, and evaluate its performance in terms of accuracy and utility. Furthermore, we have made our code available on GitHub\footnote{https://github.com/DPCodesLib/DBSec23}, enabling other researchers to reproduce our experiments and extend our work. For statistical significance, all experiments are performed 50 times and the mean values are reported as the final results.

\textbf{Environment.}
The experiments were executed on CPU: $Intel(R) Core(TM)\\ i7-11370H\ @3.30GHz$ with $Python$ version 3.7.

\textbf{Nonlinear Programming Solver.}
The solver used for solving nonlinear constrained programming is $Gurobi\ Optimizer\ version\ 10.0.1$ API for $Python$.

\textbf{Level of Temporal Correlations.}
To evaluate the performance of our post-processing method under different temporal correlations, we introduce a method to generate transition matrix in different \textit{levels}. To begin with, we default a transition matrix indicating the “strongest” correlations which contains probability 1.0 in its diagonal cells. Then, we utilize \textit{Laplacian smoothing} \cite{Sorkine04} to uniform the possibilities of $n\times n$ transition matrix $\mathcal{T}^{S}$ of `strongest' correlations. Next, let $p_{ij}$ denote the element at the $i$th row and $j$th column of $\mathcal{T}^{S}$. The uniformed possibilities $\hat{p}_{i,j}$ can be generated from \eqref{Laplacian Smoothing}, where $s$ ($0\leq s <\infty$) is a positive parameter that controls the levels of uniformity of probabilities in each row. That's, a smaller $s$ means stronger level temporal correlations. Also, We should note that, different $s$ are only comparable under the same $n$.

\begin{equation}
    \hat{p}_{i,j} = \frac{p_{ij}+s}{\sum_{j=1}^{n}(p_{ij}+s)}
    \label{Laplacian Smoothing}
\end{equation}

\subsection{Utility Analysis}
In this subsection, we conduct a utility analysis using the objective function of the nonlinear constrained programming approach described above. The objective function consists of two parts: the noise distribution and the joint distribution of query answers under temporal correlations. The key to the effectiveness of our post-processing method in achieving high utility lies in its ability to recover the correlations between data that are blurred by incremental noise added to the original query answers. For example, it enables the preservation of the correlation that `the current number of people staying at $loc1$ must equal the previous number of people staying at $loc2$' by solving the relevant nonlinear constrained programming problem. As a result, the similarity between the post-processing query answers and the original query answers is improved significantly.

\begin{figure*} [t]
	\centering
	\subfigure[MSE of Binary Counts Release]{\includegraphics[width=0.45\textwidth]{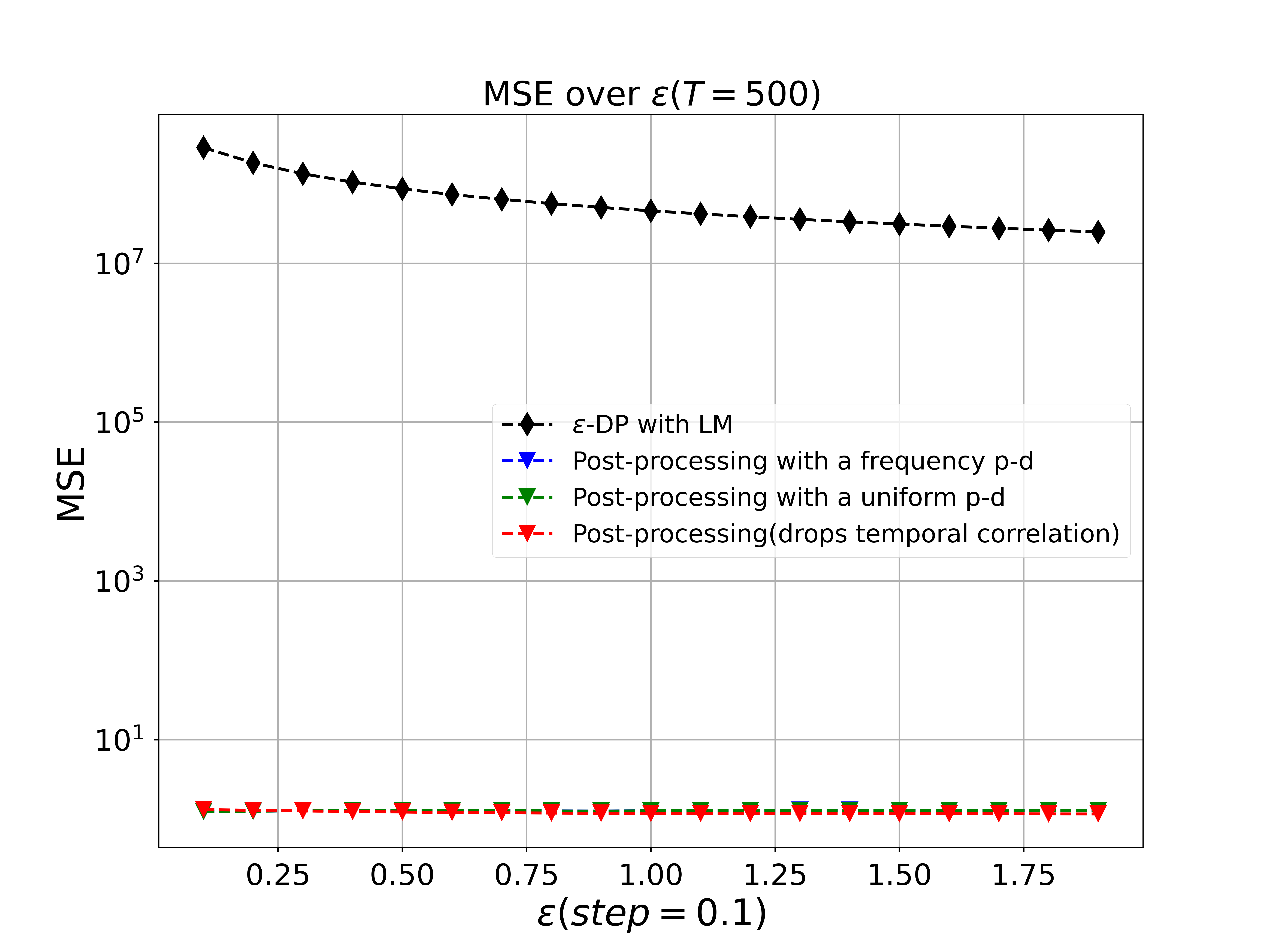}}
     \subfigure[Possibilities of Varying]{\includegraphics[width=0.45\textwidth]{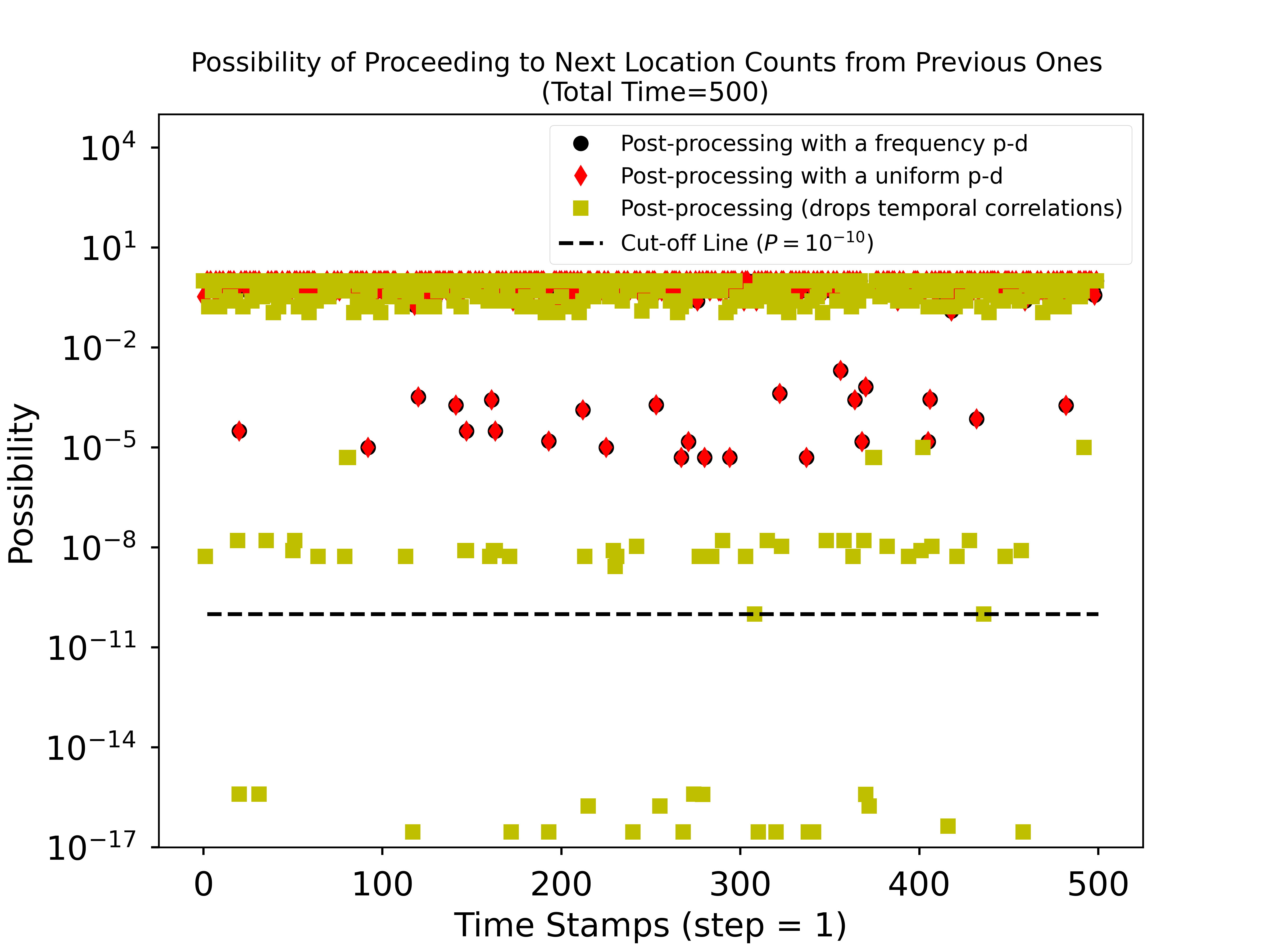}}

	\caption{ a) Scenario: Streaming Binary Counts Released under Temporal Correlations; b) Possibility of Proceeding to Current Counts from Prior Counts of Post-processing Results (under $1-$DP)}
    \label{Fig5.4} 
\end{figure*}

Moreover, we introduce $MSE$ and $Possibility$ as metrics to measure the utility of optimal counts instead of $MSE$ solely for supporting the validation of our MAP post-processing method. For instance, synthetic streaming binary counts, such that total number of locations is $n_{loc}=3$, total number of users is $n_{user}=1$, are going to be released under $\epsilon-$DP. And the temporal correlations are known to the public which can be  expressed by transition matrix $\mathcal{T}=$
$\left\{ \begin{array}{ccc}
     0.0 & 0.0 & 1.0 \\
     0.5 & 0.0 & 0.5 \\
     0.0 & 1.0 & 0.0 \\
\end{array} \right\}$ which also represents the basic temporal correlation used for generating synthetic datasets. Then, we post-process and release optimal counts using post-processing methods of MLE with a ADMM algorithm \cite{Lee15} and our MAP framework respectively illustrated by Fig.~\ref{Fig5.4}(a). The red line of Fig.~\ref{Fig5.4}(a) represents optimal results that drop temporal correlations obtained by calculating MLE problem while the greed and blue lines represent the optimal results obtained from our method of MAP framework under two different strategies. The details of them will be revealed in the following subsections.
When calculating the possibilities of achieving current counts from previous counts (shown in Fig.~\ref{Fig5.4}(b)), however, we note that many possibilities of post-processing points of dropping temporal correlations are lower than cut-off line ($10^{-10}$) which will be seen as `impossible events' if possibility is smaller than $10^{-10}$. It proves that our MAP framework is able to  preserve the probabilistic properties owned by original data, namely temporal correlations, compared with prior post-processing methods.

We will now explore the tradeoff between privacy and utility. The objective function reveals that the privacy budget $\epsilon$ is a weight parameter affecting the noises' part, but it has no impact on the correlations' part. This means that the correlations' part is dominant when the privacy budget is strict, while the noises' part replaces it when the budget is lax. As a result, the utility is always preserved under any given privacy budget, since the method always preserves known correlations when calculating the `closest' private counts.

\subsection{Synthetic Datasets}
To thoroughly examine the feasibility and effectiveness of our MAP framework and related post-processing methods, we conduct an evaluation on various synthetic datasets. This evaluation aims to provide a comprehensive understanding of the performance of our approach under different scenarios and to validate its potential for practical applications.

\textit{\textbf{MSE vs Privacy Budget $\epsilon$ or $\alpha$}}

\begin{figure*} [t]
    \centering
    \subfigure[MSE over $\epsilon$ under $\epsilon-$DP]{\includegraphics[width=0.45\textwidth]{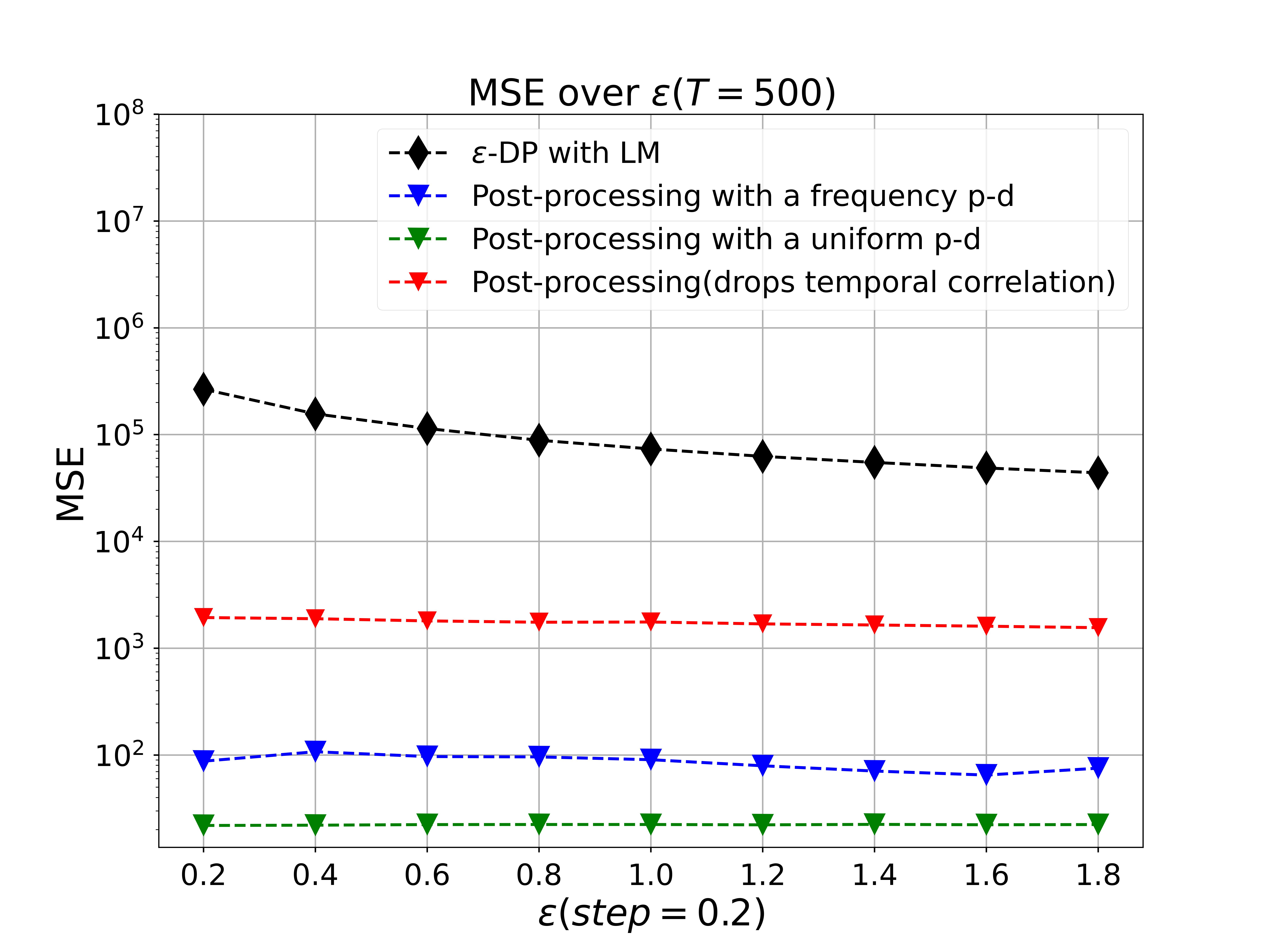}}
     \subfigure[MSE over $\alpha$ for under $\alpha-$DP$_\mathcal{T}$]{\includegraphics[width=0.45\textwidth]{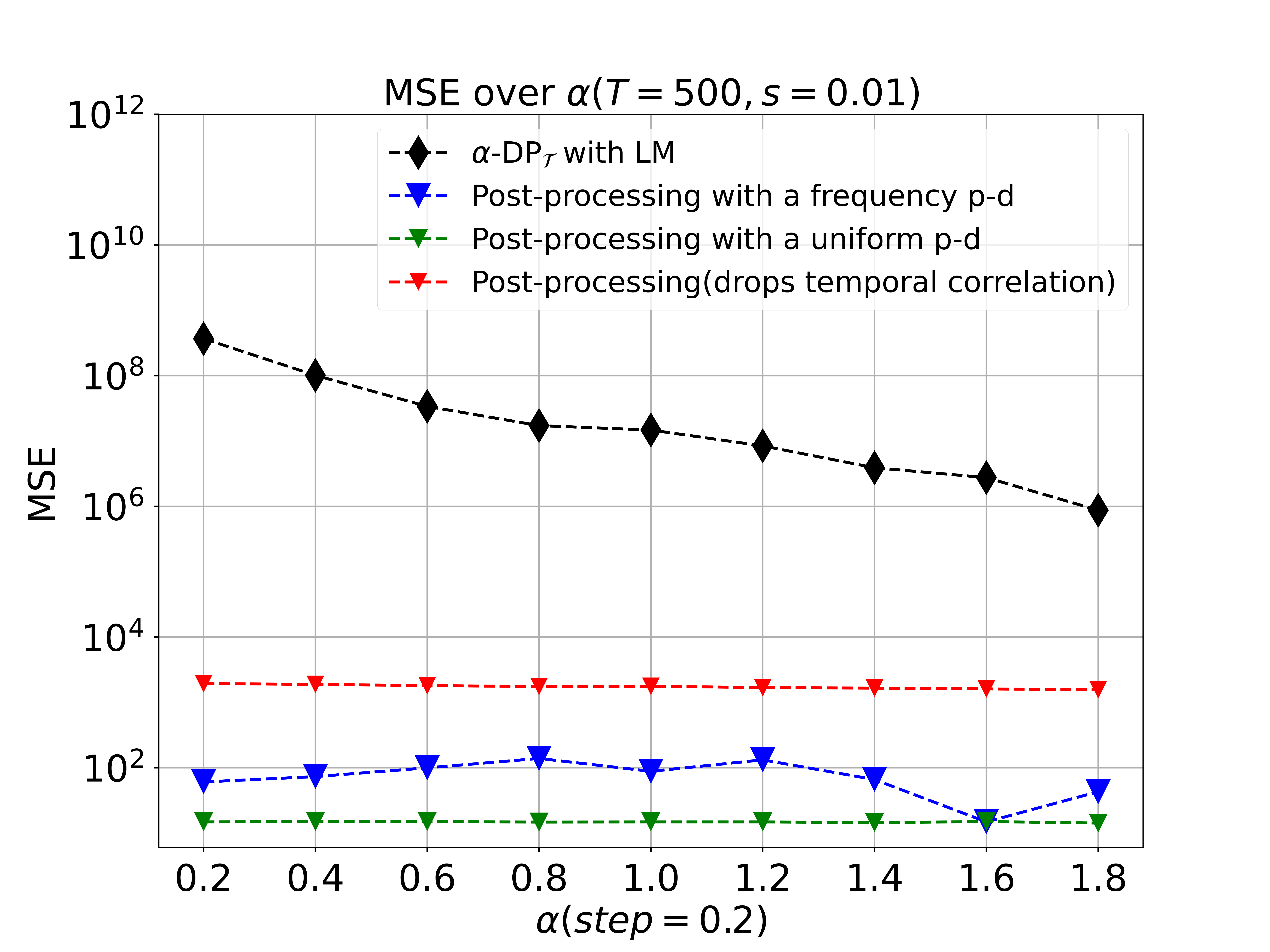}}

	\caption{ a) MSE over $\epsilon$ under $\epsilon-$DP; total release time is $T=500$, total number of users is $n_{user}=200$ and level of correlations is $s=0$. b)MSE over $\alpha$ for under $\alpha-$DP$_\mathcal{T}$; total release time is $T=500$, total number of users is $n_{user}=200$ and level of correlations is $s=0.01$.}
    \label{Fig6.1} 
\end{figure*}

Here, we compare the performance of our post-processing method by varying privacy budget $\epsilon$ or $\alpha$ from 0.2 to 2.0 (with $step=0.2$) at a given total publishing time $T$ in different mechanisms, $\epsilon -DP$ and $\alpha -DP_{\mathcal{T}}$, respectively. Note that we must choose a prior distribution(p-d) for $P^{1}$ when $t=1$, and our strategy is to use the frequency of $\tilde{R}^{1}$ or a uniform distribution to substitute for it. And the results, shown in Fig.~\ref{Fig6.1} (a) and Fig.~\ref{Fig6.1} (b), illustrate that our post-processing method significantly improves the utility and accuracy of outputs while achieving a desired privacy budget both in $\epsilon -DP$ and $\alpha -DP_{\mathcal{T}}$. 

The red line which represents prior method of MLE using ADMM only considers utilizing public knowledge of mechanisms instead of both mechanisms and correlations to boost utility of released counts. The blue line and green line are the results after our post-processing given two different policy to choose p-d. As shown in figure, MSE become smaller while increasing privacy budget $\epsilon$. And our methods perform better than MLE method by decreasing MSE nearly hundred times at any given fixed $\epsilon$.

\textit{\textbf{MSE vs Total Release Time}}

\begin{figure}[ht]
    \centering
    \includegraphics[width= 0.9\linewidth]{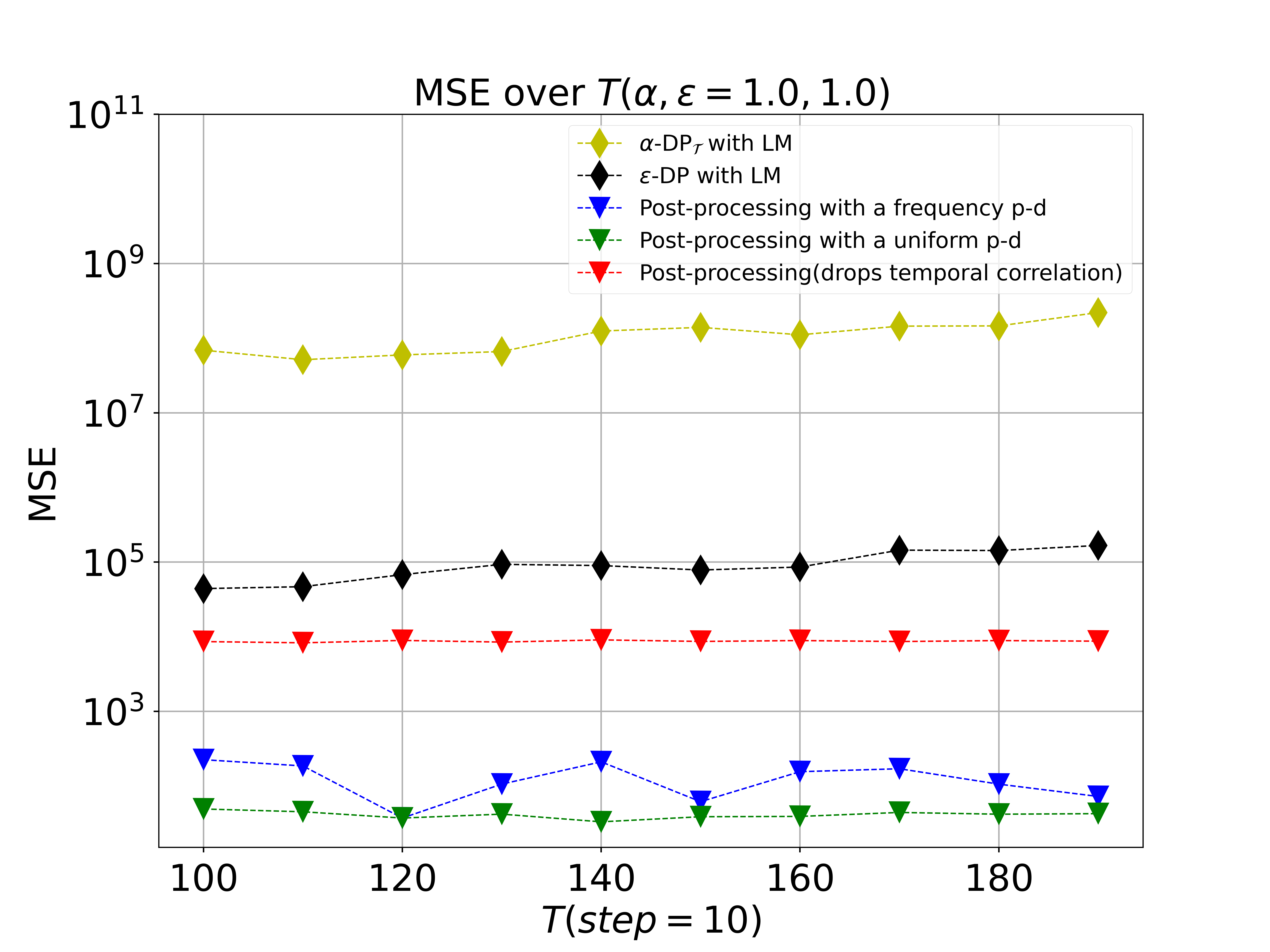}
    \caption{MSE over Total Release Time under $\alpha-$DP$_\mathcal{T}$ and $\epsilon -DP$; privacy budget is $\alpha=1.0$, $\epsilon=1.0$ and level of correlations is $s=0.01$.}
    \label{fig:MSEoverTime}
\end{figure}

We vary the total publishing time $T$ from $100$ to $200$ ($step=10$) to examine the performance of our post-processing method under both $\epsilon$-DP and $\alpha$-DP$_\mathcal{T}$, using the same methods for generating the synthetic datasets as described above. We use default privacy budgets of $\alpha, \epsilon=1.0, 1.0$.

The results of our experiments, as shown in Figure~\ref{fig:MSEoverTime}, indicate that the mean squared error (MSE) values of $\epsilon$-DP and $\alpha$-DP$_\mathcal{T}$ increase significantly as the total release time is extended. However, our post-processing method demonstrates a remarkable boost in utility, as both of its policies consistently yield lower MSE values than the method that drops temporal correlations. These findings underscore the effectiveness of our post-processing method in preserving the correlations between raw data, and the importance of considering temporal correlations when designing and evaluating different data release mechanisms.

\textit{\textbf{MSE vs Different Temporal Correlations}}

\begin{figure}[ht]
    \centering
    \includegraphics[width= 0.9\linewidth]{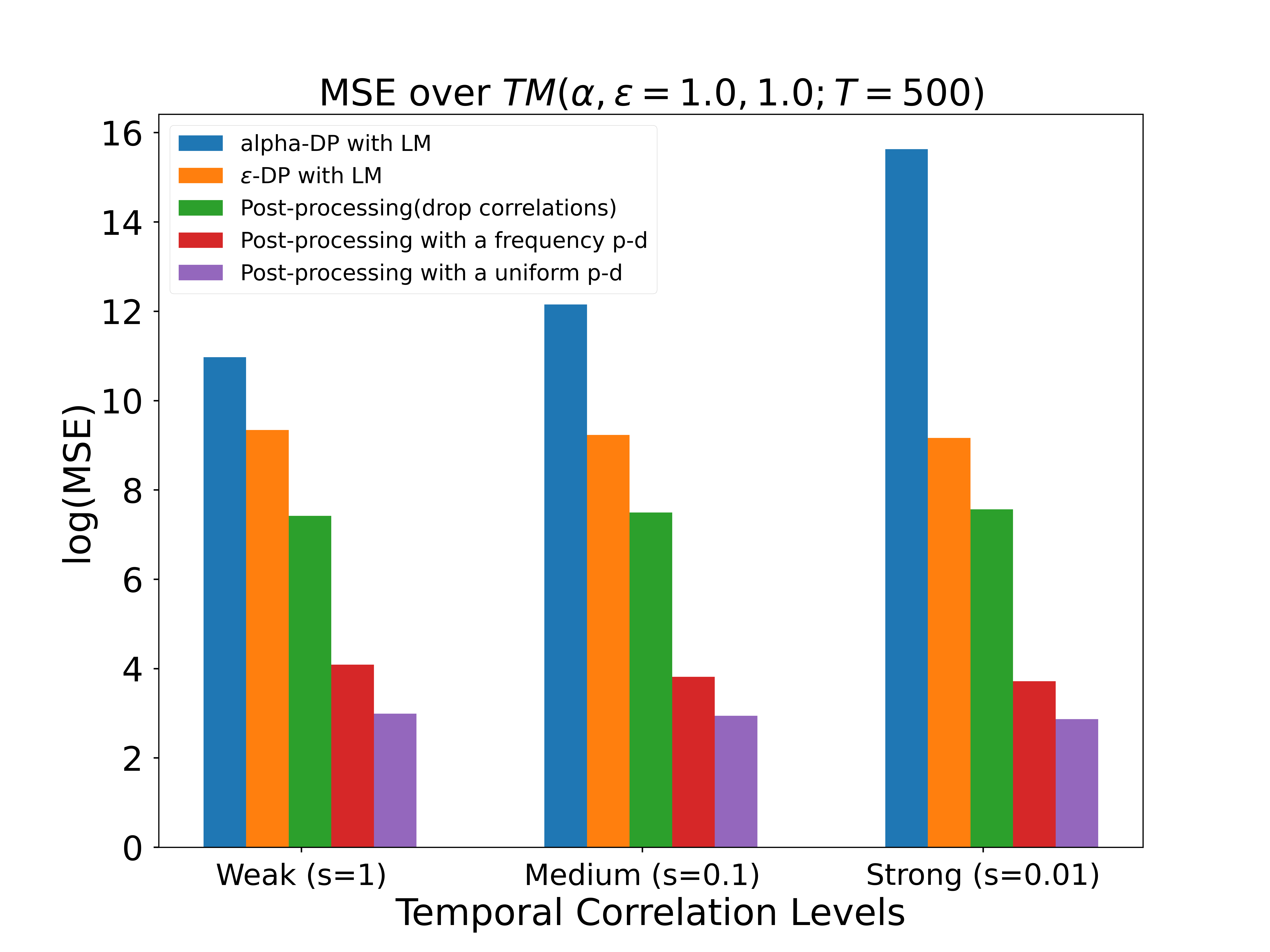}
    \caption{MSE over Different Levels of Temporal Correlations under $\alpha-$DP$_\mathcal{T}$ and $\epsilon-$DP; privacy budget is $\alpha=1.0$ and $\epsilon=1.0$, total number of users is $n_{user}=200$ and total release time is $T=500$.}
    \label{fig:MSEoverTM}
\end{figure}

In this subsection, we finally check the performance of our post-processing method upon different intensities of temporal correlations. We default the privacy budget and total publishing time as $\alpha =1.0$ and $T=500$ respectively. Note that it will have relatively higher temporal correlations if users have a higher possibility from present location to the next specific location (e.g., $Pr(l_{i}^{t}|l_{i}^{t-1})=1.0$). Therefore, we firstly generate a transition matrix  $\mathcal{T}=$
$\left\{ \begin{array}{ccc}
     0.0 & 0.0 & 1.0 \\
     0.5 & 0.0 & 0.5 \\
     0.0 & 1.0 & 0.0 \\
\end{array} \right\}$. Then, we apply \eqref{Laplacian Smoothing} to generate different level degree of correlations, weak correlations, medium correlations and strong correlations corresponding to $s=1$, $s=0.1$, $s=0.01$ respectively. 

And the results, shown in Fig.~\ref{fig:MSEoverTM}, reveal validation of this post-processing method by giving prominent improvement in accuracy. We also compare a special post-processing method that drops temporal correlations which means that the joint distribution of query results $Pr(R)$ is a constant. And the results show that the post-processing method with temporal correlations will achieve higher utility with a lower MSE.

These experiments also highlight the essential role of the MAP framework, demonstrating that correlations between raw data can significantly impact the results and cannot be disregarded in both the mechanism design and post-processing stages.


\section{Conclusion}
In this paper, we have shown that temporal correlations are often present in differential privacy data releases and proposed a MAP framework to address the post-processing problem in this context. Our experiments demonstrate the effectiveness of incorporating temporal correlations into the post-processing step, resulting in significant improvements in accuracy and utility.

Furthermore, our work suggests that the MAP framework can be a useful tool for addressing other post-processing problems involving correlated data, such as Bayesian DP and Pufferfish Privacy Mechanisms.
While our approach assumes independence between users, this may not always hold true in practice. Future work could explore how to extend our framework to address post-processing for streaming data releases under temporal correlations when users are correlated.

Overall, our work contributes to advancing the state of the art in differential privacy data releases by providing a new perspective on post-processing under temporal correlations and opens up new avenues for future research in this area.

\section*{Acknowledgments}


This work was partially supported by JST CREST JPMJCR21M2, JST SICORP JPMJSC2107, JSPS KAKENHI Grant Numbers 19H04215, 21K19767, 22H03595 and 22H00521.

Additionally, Xuyang would like to express appreciation to tutor for his meticulous instruction, as well as to his family and friend Yuchan Z. for their encouraging support in his research endeavors.

\newpage
%
%
%
\bibliographystyle{plain}
\bibliography{main}
\end{document}